\documentclass{article}
\usepackage{mrs2005,epsfig}

\newcommand{\hkpc}{h^{-1}{\rm kpc}}
\newcommand{\hmpc}{h^{-1}{\rm Mpc}}
\newcommand{\kms}{\;{\rm km}\,{\rm s}^{-1}}
\newcommand{\kmsmpc}{\kms\;{\rm Mpc}^{-1}}

\begin{document} 

\title{Building Galaxies with Simulations}

\author{Romeel Dav\'e}
\affil{University of Arizona} 
\author{Kristian Finlator}
\affil{University of Arizona} 
\author{Lars Hernquist}
\affil{Harvard University} 
\author{Neal Katz}
\affil{University of Massachussets} 
\author{Du\v{s}an Kere\v{s}}
\affil{University of Massachussets} 
\author{Casey Papovich}
\affil{University of Arizona} 
\author{David H. Weinberg}
\affil{Ohio State University} 

 \begin{abstract}
We present an overview of some of the issues surrounding current models of
galaxy formation, highlighting recent insights obtained from cosmological
hydrodynamic simulations.  Detailed examination of gas accretion processes
show a hot mode of gas cooling from near the halo's virial temperature,
and a previously underappreciated cold mode where gas flows in along
filaments on dynamical timescales, emitting its energy in line radiation.
Cold mode dominates in systems with halo masses slightly smaller than
the Milky Way and below, and hence dominates the global accretion during
the heydey of galaxy formation.  This rapid accretion path enables prompt
assembly of massive galaxies in the early universe, and results in $z\sim
4$ galaxy properties in broad agreement with observations, with the most
massive galaxies being the most rapid star formers.  Massive galaxies
today are forming stars at a much reduced rate, a trend called downsizing.
The trend of downsizing is naturally reproduced in simulations, owing
to a transition from cold mode accretion in the early growth phase to
slower hot mode accretion once their halos grow large.  However, massive
galaxies at the present epoch are still observed to have considerably
redder colors than simulations suggest, suggesting that star formation
is not sufficiently truncated in models by the transition to hot mode,
and that another process not included in current simulations is required
to suppress star formation.
\end{abstract} 
 
\section{Introduction} 

Galaxy formation is harder than it looks.  The classic papers of the
1970's outlined broad scenarios for how galaxies form in the presence
of dark matter halos (e.g. \cite{whi78}), and together with the advent
of hierarchical structure formation, the following conventional
wisdom has dominated models for galaxy formation:  Galaxies form
within biased peaks in large-scale structure, by cooling out from a
virial-temperature halo of gas onto a rotationally-supported central
disk where stars form.  The broad success of this scenario has spawned
an increasingly sophisticated industry known as semi-analytic modeling,
building on the above framework by introducing and tuning analytic
parameters to describe the sundry gastrophysical processes that govern
the appearance and evolution of galaxies.  While promising, various
semi-analytic methods have often found non-unique solutions in parameter
space that are able to equivalently match available data, resulting in
some ambiguity for the predictions of such models.  The ultimate goal of
galaxy formation theory is to understand the process thoroughly enough
that semi-analytic (or equivalently simple and intuitive) models can
successfully and uniquely reproduce the broad range of properties that
galaxies exhibit across cosmic time.  Unfortunately, it appears that
our current level of understanding leaves us far from this goal.

To better understand and characterize galaxy formation, it is useful to
employ numerical simulations of the growth of structure that dynamically
follow the processes required to form baryonic galaxies.  Such techniques
have been developing at an accelerated pace due to a combination of
Moore's Law, algorithmic improvements in modeling complex baryonic
phyics, and the advent of a standard model of cosmology \cite{spe03}.
Since galaxy formation connects scales from sub-parsec star formation
to megaparsec structure formation, simulations are unrelentingly limited
by dynamic range; hence the continuing exponential increase in hardware
computing speed have not only enabled more accurate models, but have also
spurred greater levels of insight.  Modeling baryons better involves both
developing more accurate algorithms to follow gas dynamics, as well as
understanding how to model all the physical processes that are relevant
to galaxy formation; both areas have seen significant progress in recent
years.  Finally, the concordance cosmology now favored by a variety of
observations allows the evolution of large-scale structure and dark matter
halos to be specified precisely (or precisely enough), so that efforts
in exploring parameter space can now be expended towards constraining
more uncertain baryonic processes rather than the underlying cosmology.
These improvements together have propelled simulations into becoming
an indispensible tool for studying the physics of galaxy formation and
interpreting observations of galaxies.

A successful theory of galaxy formation must be able to explain a
wide range of multiwavelength observations at a variety of epochs.
X-rays observations indicate that galaxies must have a significant impact
on their environment, because intracluster gas shows elevated entropy
levels and substantial enrichment.  Optical surveys such as SDSS and 2dF
have mapped the local galaxy distribution in exquisite detail, and deeper
probes are now finding substantial populations of galaxies out to $z\sim
6$.  Near infrared surveys, a focus of this meeting reflecting the recent
detector-driven boom in this field, have enabled detailed stellar
mass assembly studies out to $z\sim 3$, along with searches for galaxies
into the reionization epoch.  Far infrared studies have also received a large
boost from {\it Spitzer}, revealing the evolution of the dust enshrouded
universe.  Sub-millimeter bolometers have detected galaxies at a range of
redshifts thanks to fortuitous K-corrections, discovering some of the most
actively star forming galaxies in the universe.  Radio observations are
useful for studying molecular gas out to $z\sim 6$, and the promise of
21~cm mapping of the high-redshift HI distribution looms on the horizon.
Forefront work in many of these areas is represented in these proceedings.

In order to model the formation of all these objects in the correct
number and by the correct time, we have at our disposal the following:
An understanding of the growth of the dominant mass component in the
universe, an empirically tight yet vague understanding of how gas
turns into stars from the Kennicutt (1998) relation, and some really
big computers.  Can we do it?  Not surprisingly, the current answer is
no, but in this review I hope to give at least a partial view of where
we stand in the process.

Galaxy formation can be thought of as having two phases:  Gas enters
into galaxies from the intergalactic medium (``accretion"), and gas
is returned from galaxies, carrying energy and metals, back into
the intergalactic medium (``feedback").  Our understanding of both
components remains incomplete, significantly more so for the latter.
In these proceedings I will discuss recent progress using cosmological
hydrodynamic simulations towards understanding how gas gets into
galaxies, present some comparisons to observations at high redshifts,
mention the origin of galaxy downsizing and implications thereof, and
discuss how massive early type galaxies reveal some outstanding issues
that simulations must confront in the coming years.

\section{How Does Gas Get Into Galaxies?} 

\begin{figure}  
\vspace*{0.cm}  
\begin{center}
\epsfig{figure=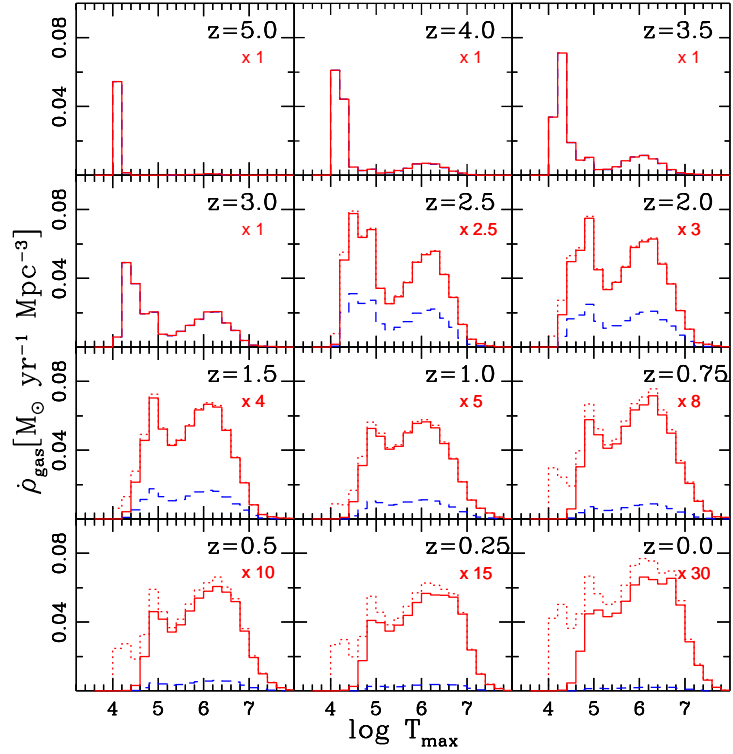,width=10cm}  
\end{center}
\vspace*{-0.5cm}  
\caption{Histograms of maximum gas temperatures achieved by particles
that turned into stars by the indicated redshift.  The bimodality of
hot and cold mode accretion is evident at all redshifts, though stronger
at high-$z$.  Cold mode dominates at early times ($z>2$), when the accretion
rates onto galaxies are the highest. From \cite{ker05}.
} 
\label{fig:diff_acc_ns}
\end{figure} 

\begin{figure}  
\vspace*{-1.cm}  
\begin{center}
\epsfig{figure=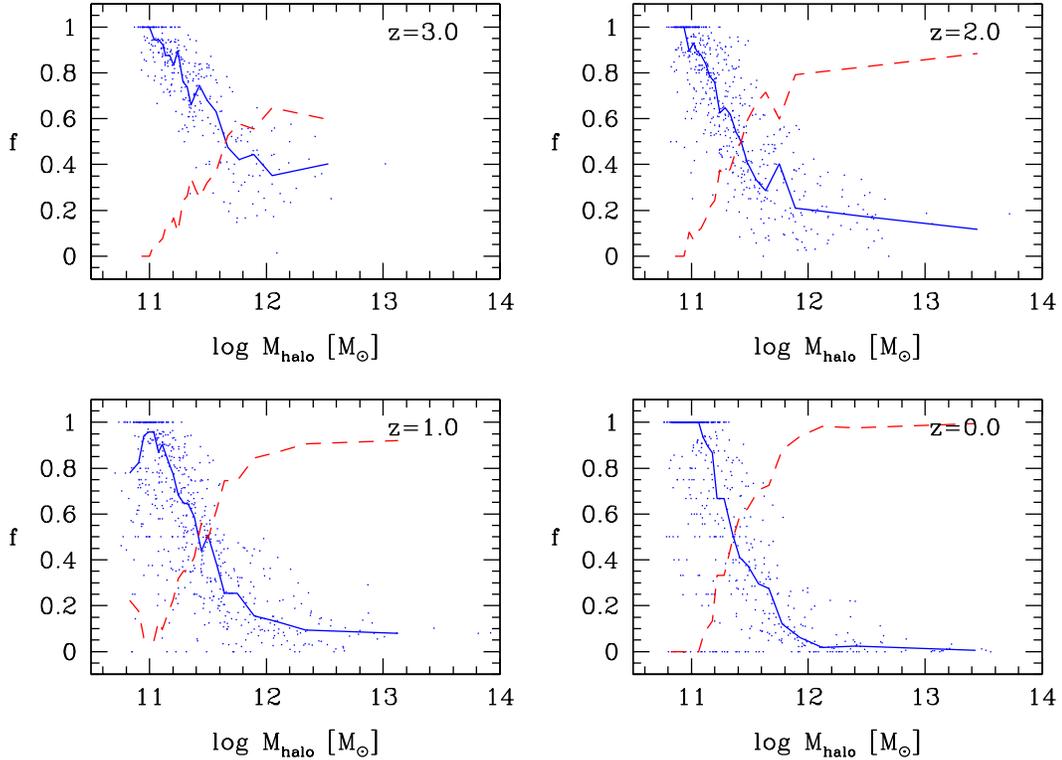,width=15cm}  
\end{center}
\vspace*{-3cm}  
\caption{Fraction $f$ of gas accreted in cold mode for galaxies at
various redshifts.  Blue line shows the median cold mode fraction, and
red line shows the median hot mode fraction.  The crossover point is
at few times $10^{11}M_\odot$, basically independent of redshift from
$z=3\rightarrow 0$. From \cite{ker05}.
} 
\label{fig:mass_ratio_halo}
\end{figure} 

Recent simulations have provided some quantitative insights into the
nature of gas accretion onto galaxies at various masses and epochs.
The White \& Rees \cite{whi78} scenario for gas accretion onto galaxies
is clearly seen to occur in simulations; hot halos form around massive
galaxies with central densities sufficiently high that a cooling flow is
established.  Indeed, later we will discuss the difficulty in preventing
such a cooling flow, as appears to be required by observations.

But this form of accretion does not appear to be the whole story, or
indeed even the majority of the story.  Kere\v{s} et al. \cite{ker05}
found that a second mode of gas accretion dominates in halos with masses
somewhat smaller than the Milky Way's and below.  To differentiate, they
dubbed accretion from cooling of virial temperature gas as ``hot mode",
and this alternate mode as ``cold mode".  Cold mode occurs when the cooling
time in the halo outskirts is sufficiently short that infall becomes
regulated instead by the dynamical time.  In cold mode, the infalling
gas temperature never approaches the virial temperature of the halo,
but instead its gravitational potential energy is radiated away as line
emission in primordial and (if present) metal atomic cooling lines.

The existence of cold mode accretion is not a new idea.  \cite{bin77}
recognized that such a mechanism should exist, and indeed all
semi-analytic models contain such a mode that gets activated whenever
the cooling time at the virial radius is shorter than the Hubble time
(or some other relevant formation timescale).  Hydrodynamic simulations
have provided additional insights into the quantitative importance of
cold mode, in particular demonstrating that owing to processes inherent in
a three-dimensional fully dynamical model, cold mode is significantly
more important than had been previously realized.

To estimate the quantitative importance of cold mode versus hot mode,
one can track the temperature history of each gas particle that ends
up as a star in a galaxy, find the maximum temperature achieved, and
determined the fraction of mass accreted at a given maximum temperature.
Such a plot (from Kere\v{s} et al.) is shown in Figure~\ref{fig:diff_acc_ns},
at a range of redshifts.  The bimodality of accretion temperatures is
evident, particularly at early times ($z>2$) when the universe is rapidly
forming stars.  Identifying the lower-temperature hump with cold mode
and the higher-temperature one with hot mode, these simulations show a
dividing temperature between modes of $\approx 2.5\times 10^5$~K.  As can
be seen, when star formation is most active in the universe at $z>2$, cold
mode is actually the dominant mechanism for gas accretion into galaxies.

Insights on the physical mechanism behind cold accretion come from
investigating the dependence of the cold mode fraction (i.e. the
fraction of star particles whose parent gas particle's temperature never
exceeded $2.5\times 10^5$~K) on galaxy and halo mass.  The clearest
trends occured with halo mass, in the sense that there appears to be
a dividing mass of $\approx 10^{11.4}M_\odot$, independent of redshift
(for $z<3$), below which accretion is predominantly cold mode, as shown
in Figure~\ref{fig:mass_ratio_halo}.  This value exceeds by a factor of
a few the mass threshold obtained by simply equating the cooling time to
the Hubble time at the virial radius.  An independent investigation by
\cite{bir03} using a one-dimensional galaxy formation model including
primordial cooling and cosmological accretion found a threshold for
stable accretion shock formation of below $10^{11}M_\odot$, comparable
to the value from a simple virial radius cooling time argument.  Hence it
is the three-dimensional geometry of infall, particularly enhancement
of density along filaments~\cite{kat02}, that is responsible for the
increased importance of cold mode accretion.

The exact temperature of the division between hot and cold accretion
depends on a variety of parameters.  The investigations above only
included primordial cooling, whereas if metal line cooling is included the
dividing mass will presumably rise.  The exact value is also sensitive to
the ratio of $\Omega_m/\Omega_b$, which was assumed to be 7.5 in Kere\v{s}
et al.; a more canonical concordance value of 6 would likewise result
in a higher dividing mass.  It is probable therefore that cold mode is
even more important overall than suggested in Kere\v{s} et al.  In any case,
more significant than the precise values are the qualitative notions that
(1) Cold mode dominates in smaller halos, and hence is more important
at earlier epochs; (2) Cold mode dominates in galaxies with masses up to
$M_*$ and higher at redshifts when the universe is most rapidly forming
stars; and (3) Cold mode arises from the lack of virial shock pressure
support in the presence of radiative cooling.

\section{An Early Epoch of Star Formation}

One consequence of having a rapid mode of accretion being effective at
high redshift is that it enables massive galaxies to form their stars
quickly in the early universe.  Such an early epoch of star formation
has been inferred from the observed stellar populations of early type
galaxies out to $z\sim 2$.  The galaxy population in the early universe
therefore presents an interesting test of the overall simulation picture
of stellar mass assembly and gas accretion in galaxies.

\begin{figure}  
\vspace*{0cm}  
\begin{center}
\epsfig{figure=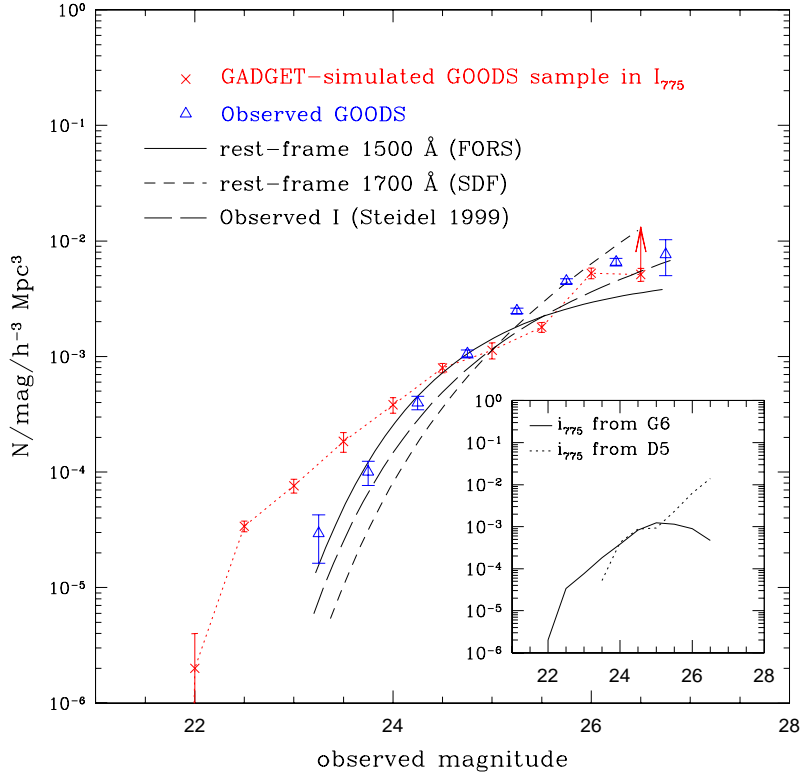,width=11cm}  
\end{center}
\vspace*{-0.5cm}  
\caption{ The simulated rest-frame UV LF of B-dropouts compared with
several published curves and the GOODS data. Red crosses indicate
the simulated LF in observed $i_{775}$.  The GOODS B-dropout sample 
is indicated by blue triangles.  The simulated LFs are in broad agreement
with available constraints, with a possible excess at the bright end.
{\it Inset:} The LFs of the D5 and G6 simulations, which were combined
to produce the full curve.  From \cite{fin05}.
} 
\label{fig:lf_goods}
\end{figure} 

To study this, Finlator et al. \cite{fin05} examined the population of
galaxies in Gadget-2 \cite{spr03} simulations of $z=4$ galaxies, as would
be seen in the color selected B-dropout sample of the Great Observatories
Origins Deep Survey (GOODS).  By calculating the photometric properties
of galaxies using the star formation histories convolved with the Bruzual
\& Charlot (2003) population synthesis code, we predicted the luminosity
function as would be observed using a Lyman break selection technique at
$z\approx 4$.  We also included dust reddening using a novel prescription
calibrated from the local metallicity-extinction relation from SDSS.

The results are shown in Figure~\ref{fig:lf_goods}.  The agreement is
quite good around $L_*$, but it appears the simulations produce too many
brighter galaxies.  Recent work from the VVDS Survey \cite{lef05} that
does not use color selection suggests that many bright galaxies are missed
by the color criteria typically used for Lyman break selection.  Hence it
remains to be seen if the excess actually indicates a substantial failing
of the model, or merely an improper reckoning of selection effects.

\begin{figure}  
\vspace*{0.cm}  
\begin{center}
\epsfig{figure=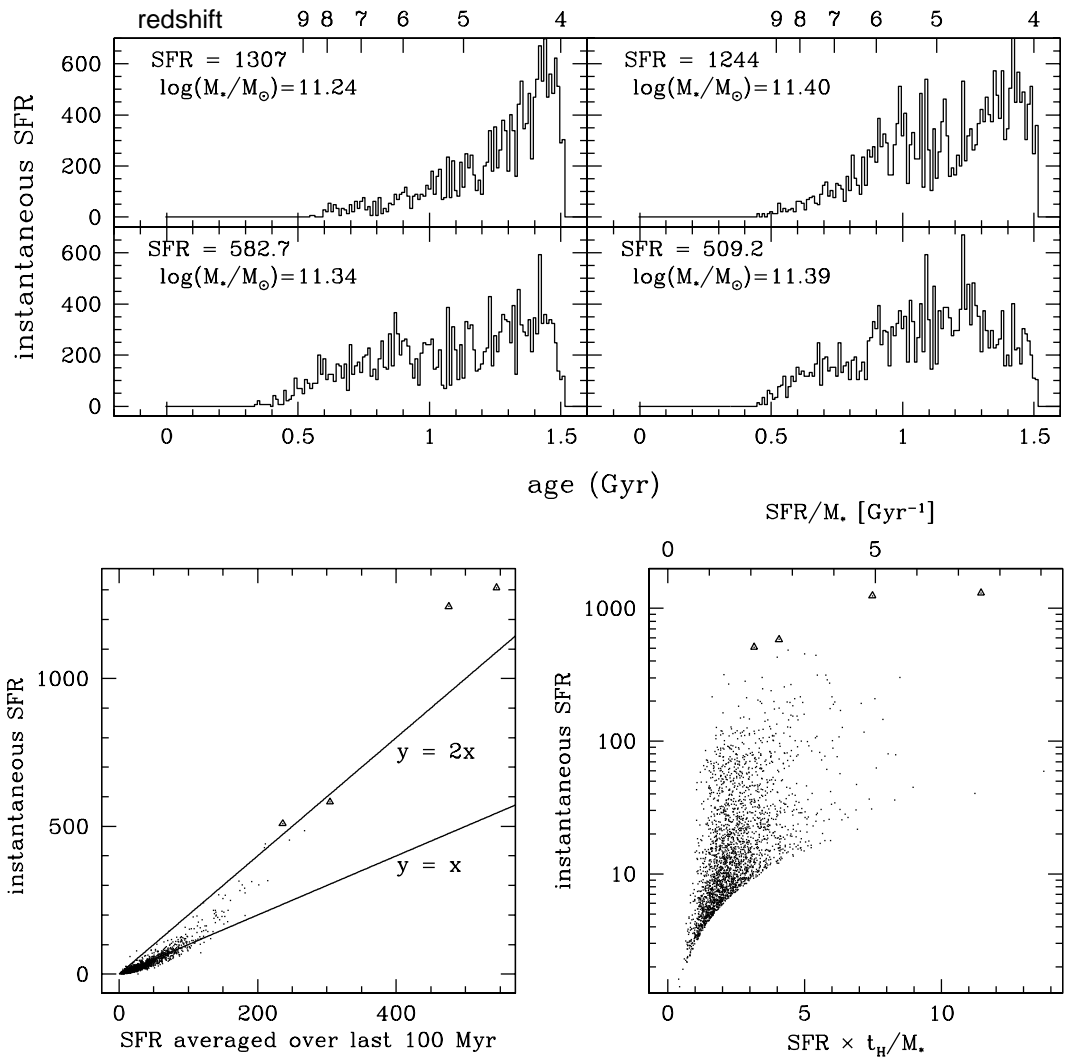,width=15cm}  
\end{center}
\vspace*{-1cm}  
\caption{Top panel shows star formation histories of the four most
rapidly star forming galaxies in the simulated samples.  The instantaneous
star formation rates and stellar masses at $z=4$ are given in the top
left.  Bottom left panel shows instantaneous versus time-averaged star
formation rate for the GOODS-selected galaxies in the G6 simulation,
with the four most rapid star formers marked by triangles.  Bottom
right panel shows instantaneous star formation rate versus birthrate.
Lower envelope arises from our $10^{9.64} M_{\odot}$ mass resolution cut.
These massive, rapid star formers are experiencing at most a mild burst
of star formation and may be identifiable as high-redshift submillimeter
galaxies. From \cite{fin05}.
} 
\label{fig:biggals}
\end{figure} 

Another possibility is that the bright galaxies are actually much dustier
than is assumed by the reddening prescription used (so that they would not
be seen in UV-selected samples), and in fact that such galaxies should
instead be associated with sub-millimeter sources detected by SCUBA.
Figure~\ref{fig:biggals} shows the instantaneous star formation rates of
these bright galaxies, which in some cases exceeds 1000~$M_\odot$/yr,
similar to that inferred for SCUBA galaxies.  The star formation
histories of the most rapid star formers are shown in the top panel
of Figure~\ref{fig:biggals}, and show a time-averaged rate of $\sim
500 M_\odot/yr$ in the recent past.  This means that these galaxies are
undergoing only a mild burst of star formation, by a factor of a couple,
over their ``quiescent" state.  This is shown more clearly the lower left
panel of Figure~\ref{fig:biggals}, plotting the instantaneous versus
time-averaged star formation rates for all galaxies in the G6 run.
Visual inspection indicates that these rapid star formers live at the
centers of protoclusters, and exhibit a disturbed morphology due to the
local high density of galaxies.

Overall, the latest observations of galaxy properties at $z\sim 4$ are
broadly reproduced in simulations.  This is reassuring, both because
it means that gas accretion processes are probably being modeled in
a reasonable manner, and also because it appears that there are no
physical processes that are unaccounted for in these models that have
a large impact on the galaxy population.  As we will see in the next
section, this statement is not true at lower redshifts.

\section{What Stops Gas From Getting Into Galaxies?} 

\begin{figure}  
\vspace*{0cm}  
\begin{center}
\epsfig{figure=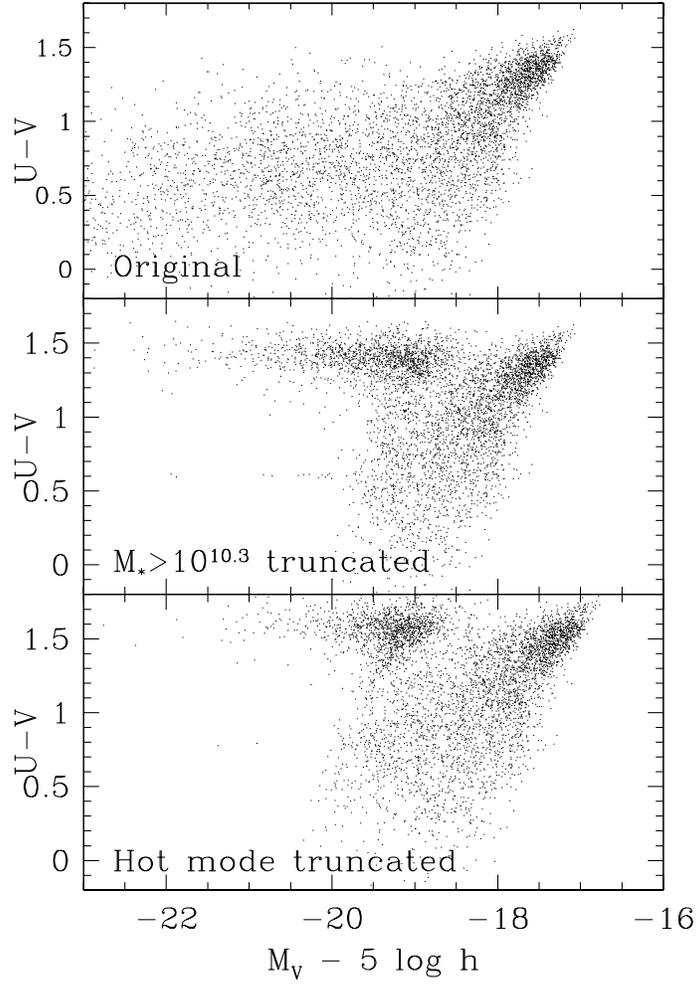,width=12cm}  
\end{center}
\vspace*{0cm}  
\caption{$U-V$ vs. $V$ color-magnitude diagrams (CMDs) from the G6
Gadget-2 simulation \cite{fin05}.  Top panel shows the original CMD
at $z=0$, obtained by computing magnitudes from the star formation
history as given by the original simulation.  Middle panel shows the
result of modifying the star formation history (post facto) so that all
star formation events are removed once the galaxy mass exceeds $2\times
10^{10} M_\odot$; this moves all massive galaxies onto the red sequence,
but unfortunately also removes all large blue galaxies.  The bottom panel
shows the result on removing all star formation events that occured due
to hot mode accretion; this looks somewhat more promising, although 
some odd features remain.
} 
\label{fig:redseq}
\end{figure} 

While agreement at high redshifts is comforting, it is obtained in part
because the data at those epochs is not of the same quality and quantity
as is available locally.  Galaxy observations in the local universe
cover a wide range of objects at a wide range of wavelengths, and there
are many outstanding issues that remain to be resolved theoretically.
Here we focus on massive early-type systems that should, in principle,
be among the easiest and simplest systems to model.  As we consider the
various observations of massive ellipticals out to $z\sim 1$, one common
feature unifies these objects:  They are virtually all ``red and dead".

This feature is best illustrated by the red sequence in a color-magnitude
diagram.  The uniformly old stellar populations together with a
mass-metallicity correlation produce a tight locus in color-magnitude
space in which more luminous galaxies are redder.  The tightness of
this relation is indicative of the lack of recent start formation.
It is this feature that is among the most difficult to understand about
galaxy formation.

In simulations, massive galaxies sit at the bottom of large potential
wells, where intergalactic gas is constantly infalling.  This is the
classic scenario for hot mode accretion, where the central gas density
is high enough for cooling times to be rapid, resulting in a substantial
amount of gas cooling and subsequent star formation.  Shutting this gas
supply off requires not only tremendous amounts of feedback energy, but also
requires that that energy be distributed quasi-spherically among the
gas elements in the cooling flow region.  What physical process could
be responsible for turning off gas accretion in massive galaxies?

The current wisdom is to invoke AGN as the heating mechanism.
This idea has a number of positives, in that there is plenty of energy
in the central engine, it can be generated with a very small amount of
accompanying gas infall or star formation, and there is a tight observed
relation between central black hole and bulge masses.  Indeed, Kauffmann in
these proceedings has assembled a compelling set of observations showing
that AGN are responsible for moving galaxies onto the red sequence.
Simulations of galaxy mergers with AGN feedback \cite{dim05} reproduce
many observed properties of quasars \cite{hop05}, and succeed in heating
gas sufficiently to prevent star formation.  All in all, there is
accumulating evidence pointing towards AGN shutting off star formation,
although much of it remains circumstantial.  The main remaining difficulty
is trying to understand how energy from the black hole couples to the gas
in such a uniform and widespread manner so as to shut off cooling flows.

Whatever the mechanism, it is clear that star formation must be truncated,
rapidly and at an early epoch, in massive galaxies.  One approach is
to identify what physical parameter governs when a galaxy will shut off
its accretion.  Is it the galaxy mass?  Is it hot mode accretion?  We can
roughly test these ideas by turning off star formation in massive systems
by post-processing their simulated star formation histories.

To do so, we take the G6 Gadget-2 simulation of \cite{spr03} (described in
\cite{fin05}), identify all galaxies at $z=0$, and compute each galaxy's
$U$ and $V$ absolute magnitudes.  The top panel of Figure~\ref{fig:redseq}
shows the resulting simulated color-magnitude diagram.  As can be seen,
there is no red sequence evident at $U-V\approx 1.5$ as observed for real
galaxies, and more massive galaxies do not become progressively redder.
This is because of residual star formation in massive systems due to
continued cooling and infall.

We now take each galaxy, track its growth back in time, and eliminate all
star formation once its progenitor's stellar mass exceed $2\times 10^{10}
M_\odot$.  The resulting star formation histories are then used to
recompute the CMD, shown in the middle panel of Figure~\ref{fig:redseq}.
This approach results in a decent looking red sequence (although it does
not have the proper slope, but this may be due to metallicity effects not
included here).  However, it is now missing another important population:
Massive blue spiral galaxies, such as our own Milky Way!  This simple stellar
mass cut apparently truncates star formation in all galaxies, rather
than only the early types.

Another idea, suggested by \cite{ker05} and \cite{dek04}, is that whatever
mechanism is responsible for stopping cooling flows preferentially does
so by keeping virial temperature gas hot, and not allowing it to cool.
Again, we can examine this in our simulation by tracking the temperature
history of each particle, and removing all star formation events if the
parent gas particle's temperature had at any time exceeded $2.5\times
10^5$~K, i.e. if it is a hot mode accretion event.  The resulting CMD
is shown in the bottom panel.  This scenario is at best only slightly
more successful, as once again bright blue galaxies have mostly been
eliminated, and larger galaxies are not particularly redder.  It seems
that while the hot/cold accretion scenario may play some role in the
truncation of star formation in massive systems, it does not appear to
be the governing factor.

Semi-analytic models of \cite{som05} and \cite{cro05} have had some
success producing a red sequence, in the former case by truncating star
formation when the bulge mass (not total mass) exceeds a certain value,
and in the latter case by shutting off gas accretion onto satellite
galaxies and including a ``radio mode" of feedback tied to AGN growth.
These parameterizations serve as useful guides for constraining the
phenomena that may be responsible.

\section{Birthrates and Downsizing}

\begin{figure}  
\vspace*{0cm}  
\begin{center}
\epsfig{figure=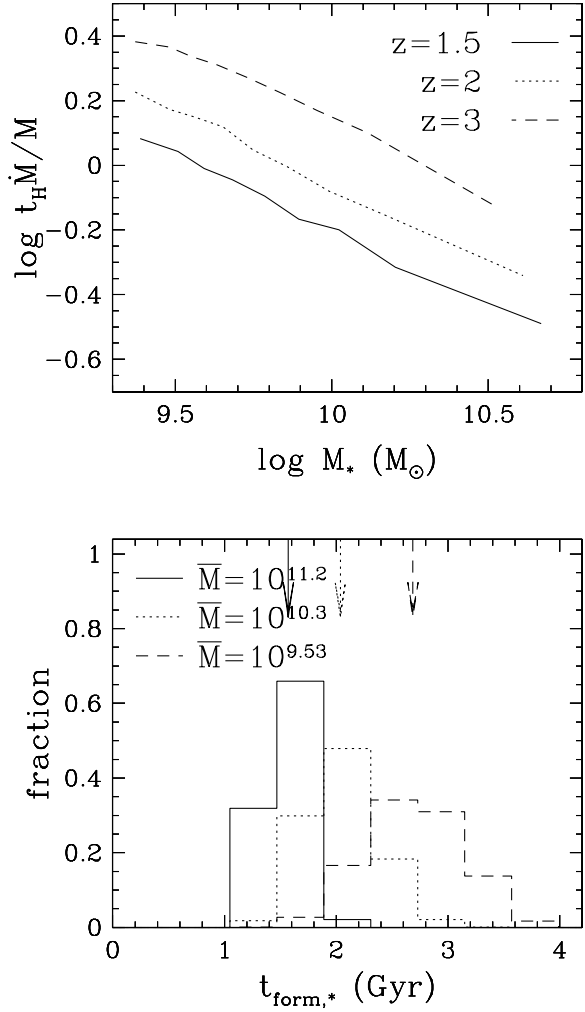,width=12cm}  
\end{center}
\vspace*{0cm}  
\caption{Downsizing in Gadget-2 simulations of galaxy formation with
superwinds turned off (cf. \cite{spr03}).  Top panel shows
the median birthrate as a function of galaxy stellar mass at 
$z=3$, 2, and 1.5.  The birthrates decrease with time as the global
star formation rate decreases.  At all epochs shown, massive galaxies are
forming stars more slowly relative to their stellar mass, in accord with
the downsizing phenomenon.  Bottom panel shows a histogram of the median
star formation times of galaxies at $z=1.5$, in three galaxy mass bins.  
Large galaxies form their stars earlier, and therefore have older stellar
populations.  Hence in hierarchical galaxy formation models, the star
formation time and halo formation time are in general inversely correlated.
While these qualitative trends of downsizing are in agreement with
observations, massive galaxies are still forming stars too rapidly
to place them on the red sequence, hence an additional physical process
is still required to truncate star formation in these systems.
} 
\label{fig:birthrate}
\end{figure} 

The formation of a red sequence is part of a larger trend in galaxy
formation known as downsizing \cite{cow96}.  Downsizing states that massive
galaxies form their star earlier and are comparatively quiescent now,
while smaller galaxies have formed (or are forming) most of their stars
today.  This trend can be summarized using the birthrate parameter,
which measure the current star formation rate normalized to the past
averaged star formation rate.

To study birthrates, we run a Gadget-2 simulation having $32\hmpc$
box size, $2.5\hkpc$ softening, and $2\times 256^3$ particles,
turning off superwind feedback so we can more cleanly study the basic
processes of accretion and star formation as a function of galaxy mass.
Figure~\ref{fig:birthrate} shows the birthrates from this simulation
at $z=3$, 2, and 1.5, where the past averaged star formation rate has
been computed over a Hubble time.  A galaxy with a birthrate parameter
of unity, for example, would double its stellar mass in another Hubble time
(assuming no merging).

Downsizing reflects the trend that massive galaxies have lower birthrates
at the present time than less massive galaxies.  Such a trend is evident
in the top panel of Figure~\ref{fig:birthrate}, at every epoch shown.
Another way to characterize this plot is to say that the typical
mass of actively star forming galaxies (e.g. those with birthrate
parameters exceeding unity) reduces with time.  Thus the observed
trend of downsizing is in accord with predictions of hierarchical galaxy
formation models.

The sense of the trend can be easily understood in terms of hot and cold
mode accretion:  Big galaxies reside in large density fluctuations
that collapse first in the early universe, and start forming stars
vigorously owing to rapid cold mode accretion.  As they grow in size, a
virial shock is able to form which slows down the accretion, leading to a
higher birthrate in the past than at the present day-- i.e., downsizing.
Smaller galaxies continue to form stars vigorously even today as they
collapse later and continue to be dominated by cold mode accretion.
Although the transition from hot mode to cold mode is fairly sharp in
halo mass, when viewed as a function of galaxy stellar mass it is 
much more gradual, which is why no obvious feature exists in the
birthrate plot marking the transition between hot and cold mode.

Another statement of downsizing is that the formation time of
stars is older in more massive galaxies.  The bottom panel of
Figure~\ref{fig:birthrate} shows a histogram of the median formation
time of stars in galaxies subdivided into three mass bins.  This plot
is made at $z=1.5$, when the universe is 4.2~Gyr old ($\Omega=0.3$,
$H_0=70\kmsmpc$, $\Lambda$CDM).  This shows that large galaxies typically
form their stars at earlier times.  Here, galaxies with $M_*\approx
2\times 10^{11}$ have a median star formation redshift of $z\approx 4$
(1.5~Gyr), while for galaxies with $M_*\approx 3\times 10^9 M_\odot$
it is around $z\approx 2.5$.  Note that the median star formation time
is {\it inversely} correlated with the median halo formation time,
since large halos assemble later in hierarchical models.  Hence the
``anti-hierarchical" appearance of downsizing is in fact a natural
outcome of galaxy assembly within hierarchical structure formation models.

Unfortunately, the qualitative trend of downsizing is only part of the
story.  Quantitatively, it appears that cosmological simulations cannot
self-consistently reproduce the {\it strength} of downsizing in massive
galaxies.  While massive galaxies form stars more slowly relative to their
stellar mass than smaller ones, they are still forming some stars, whereas
observations indicate they don't form stars at all (though see new GALEX
results in these proceedings showing residual star formation in $\sim
20\%$ of present-day ellipticals).  So some physics is still missing.

A topical implication of downsizing is that massive galaxies must grow
substantially by so-called dry merging, i.e. mergers of predominantly
stellar systems.  The argument for this simply goes that if today's
massive systems form their stars early while their masses are still small,
then in order to grow into massive systems we see today they must merge
with other large systems who must have also formed their stars early.
These mergers will then be predominantly stellar.  Indeed, there is
growing evidence for ubiquitous dry merging locally (see van Dokkum,
these proceedings), and out to $z\sim 0.7$ \cite{bel05}.

In summary, the trend of downsizing is naturally expected in a
hierarchical structure formation scenario, arising from the interplay of
biased galaxy formation and gas accretion processes.  This trend implies
that dry merging is an important growth path for massive galaxies.
However, the strength of the downsizing in massive galaxies seems to
be underestimated in current models, requiring a new form of feedback
possibly associated with black hole growth to sufficiently truncate star
formation in today's ellipticals.

\section{Conclusions} 

Galaxy formation remains an unsolved problem.  Hydrodynamic simulations
have enabled many advances in understanding the accretion processes
by which galaxies obtain their fuel for growth, including a newfound
appreciation for the importance of cold mode accretion.  Observations of
high-redshift galaxies provide a critical test for models, and current
comparisons suggest that simulations are doing a reasonable job producing
big galaxies in their youth.  A critical test appears to be the formation
of red sequence galaxies at low redshifts, something that is currently
pointing towards new feedback processes such as AGN heating.  The coming
years will see continued dramatic advances in observations, and it is
incumbent upon theoretical models to keep pace.  A solid foundation is
in place connecting primordial fluctuations and the objects we see today,
but the details remain a work in progress.

\acknowledgements{ We thank V. Le Brun and all the organizers for
a fabulous meeting.  We also thank V. Springel for providing us with
the Gadget-2 code used throughout this work.  Some of the simulations 
presented here were done at the National Center for Supercomputing
Applications (NCSA).
}

\end{document}